\documentclass[showpacs,pra,aps,twocolumn,floatfix]{revtex4}
\usepackage{graphicx,float,amsmath,amssymb}

\begin{document}

\newcommand{\vt}{{\vec \theta}}

\title{Localization by bichromatic potentials versus Anderson localization}
\author{Mathias Albert and Patricio Leboeuf}
\affiliation{Laboratoire de Physique Th\'eorique et Mod\`eles Statistiques, CNRS, Universit\'e Paris Sud, UMR8626, 91405 Orsay Cedex, France}

\begin{abstract}
The one-dimensional propagation of waves in a bichromatic potential may be modeled
by the Aubry-Andr\'e Hamiltonian. The latter presents a delocalization-localization transition, 
which has been observed in recent experiments using ultracold atoms or light. It is shown here that, 
in contrast to Anderson localization, this transition has a classical origin, namely the localization 
mechanism is not due to a quantum suppression of a classically allowed transport process. Explicit comparisons 
with the Anderson model, as well as with experiments, are done.
\end{abstract}


\maketitle


\section{Introduction}

It is by now well established that disorder may lead to the absence of diffusion
in the propagation of linear waves (Anderson localization). The phenomenology of 
Anderson localization strongly depends on dimensionality. In one dimension (1D) and
for uncorrelated disorder, it has been shown that all wave functions are exponentially 
localized. This implies that the transmission of a wavepacket incident on a disordered 
region of length $L$ decreases as $~ \exp ( -L/L_{loc} )$, where $L_{loc} \ll L$ is 
the localization length. This phenomenon is valid for arbitrary energies, even when the 
energy $E$ of the incident wavepacket is much higher than the typical amplitude of the 
disordered potential. In this respect, Anderson localization is a strongly non-classical 
phenomenon.

Another class of interesting potentials are quasiperiodic potentials. They arise,
typically, from the superposition of two periodic lattices with incommensurate periods.
Part of their interest is that they provide an example of intermediate
potentials between ordered (e.g. periodic) and disordered potentials: the system has 
neither translational symmetry nor true disorder. In particular, the model we consider 
here (the Aubry-Andr\'e Hamiltonian) is known to present a
delocalization-localization transition as a parameter is varied; wavefunctions change
from extended to exponentially localized \cite{aubry}. 

In recent experiments both random \cite{aspect} and quasiperiodic potentials \cite{inguscio} 
have been used to directly observe the localization of matter waves. The expansion of 
non-interacting Bose-Einstein condensates through one-dimensional quasiperiodic potentials 
was shown to clearly display a delocalized-localized crossover (with a shift, however, with 
respect to the predicted critical value of the parameter) \cite{inguscio}. A similar transition 
has been recently observed with light propagating in quasiperiodic photonic lattices \cite{lahini}. 
Our purpose here is to provide a semiclassical analysis of the experiments and, more generally, 
of the Aubry-Andr\'e Hamiltonian, which allows for a simple and lucid picture of the underlying 
physical phenomena (for related semiclassical analysis see e.g. 
\cite{azbel,suslov,thouless,wilkinson,leb90,leb89}). In particular, 
and in contrast to what is quite often claimed, the analysis shows that there are deep fundamental 
differences between the Anderson localization and the localization in quasiperiodic potentials, 
although both present exponential localization of the wavefunctions.

As already pointed out above, Anderson localization is a strongly non-classical 
phenomenon, where a quantum particle of energy $E$ incident on a 1D disordered region of 
length $L\rightarrow \infty$ is reflected with probability one, even though its incident energy is much 
higher than the maximum height of the disordered potential. Thus, in Anderson localization 
a quantum particle is perfectly reflected, whereas its classical counterpart is perfectly 
transmitted. In this sense, Anderson localization is a purely wave-mechanical effect. In 
contrast, as shown below, localization in quasiperiodic potentials is different. It is a purely 
classical effect, not due to any interference or wave-like dynamics. Indeed, already at the classical level
the trajectories display a delocalized-localized crossover. In this integrable model the wavefunctions 
can directly be associated to some specific classical orbits, around which they are 
exponentially concentrated. As a parameter is varied, they simply follow the classically driven 
transition. Quantum mechanically, there is thus no suppression of a classical 
diffusion process. It is our purpose to stress these aspects of the model, and to clarify 
the connections with recent experiments.

\section{The model}

The Hamiltonian we are interested in is of the tight-binding form, with 
nearest-neighbor hopping between discrete position sites $q_n$ and a superimposed quasiperiodic potential
\begin{equation}
  \begin{array}{lll}
   \hat H &=& \displaystyle J\sum_n \left( | q_{n+1} \rangle \langle q_n| + |q_n\rangle \langle q_{n+1}|
               \right) \\
     &  + & \displaystyle W \sum_n \cos(2\pi\beta n)|q_n\rangle \langle q_n| \, ,
 \end{array}
  \label{htb}
\end{equation}
where $\beta$ is the commensurability parameter, $|q_n \rangle$ is a localized Wannier state on site $n$ of the lattice, and $J$ and $W$ are parameters that control the amplitude of the hopping transitions and of the quasiperiodic potential, respectively. Since the Hamiltonian is invariant under the increase of $\beta$ by an arbitrary integer, we can restrict to $0<\beta \leq 1$. This Hamiltonian is known as the Aubry-Andr\'e or Harper Hamiltonian in the litterature \cite{aubry,harper}. There are two well known physical problems effectively described by Eq.~(\ref{htb}). The first one is the motion of electrons in two dimensions for a periodic potential in the presence of a magnetic field applied perpendicular to the plane, when interband transitions are neglected \cite{ashcroft}. In this case $\beta$ is related to the ratio of the area of the magnetic flux quantum to the unit cell in coordinate space. The second one, directly related to the experiments described here, is the 1D motion of particles in the presence of two superimposed periodic potentials, the main one of period $\lambda_1$ that determines the position of the discrete lattice points $q_n$, and the perturbing one of period $\lambda_2$; in this case $\beta=\lambda_1/\lambda_2$. The connection between the latter problem and Eq.~(\ref{htb}) is explicitly described in the next section and in Appendix A.

Expressing the hopping between nearest-neighbor sites of the main lattice in terms of the translation operator
$\exp (-i \lambda_1 \hat p /\hbar)$ ($\hat q$ and $\hat p$ are the usual position and momentum operators, respectively), one can transform Eq.~(\ref{htb}) into the more symmetric form
\begin{equation}
  \hat H=2J\cos(2\pi\hat p/P)+W \cos(2\pi\hat q/Q)\, ,
  \label{hh}
\end{equation}
where $P=2\pi\hbar/\lambda_1$ and $Q=\lambda_2$. The classical analog of this Hamiltonian is obtained by replacing the operators by $c$-numbers
\begin{equation}
  \mathcal H (q,p) = 2J\cos(2\pi p/P)+W \cos(2\pi q/Q)\,.
  \label{hc}
\end{equation}
This classical Hamiltonian is periodic in both position and momentum. Therefore its study can be restricted to a single cell of size $(Q,P)$. Note that $\mathcal H (q,p)$ contains, through the momentum scale factor $P=h/\lambda_1$, quantum Planck's constant. Its validity and applicability, as well as the apparent paradox that it contains a quantum scale, will be discussed below. 

\begin{figure}
  \includegraphics[width=\linewidth]{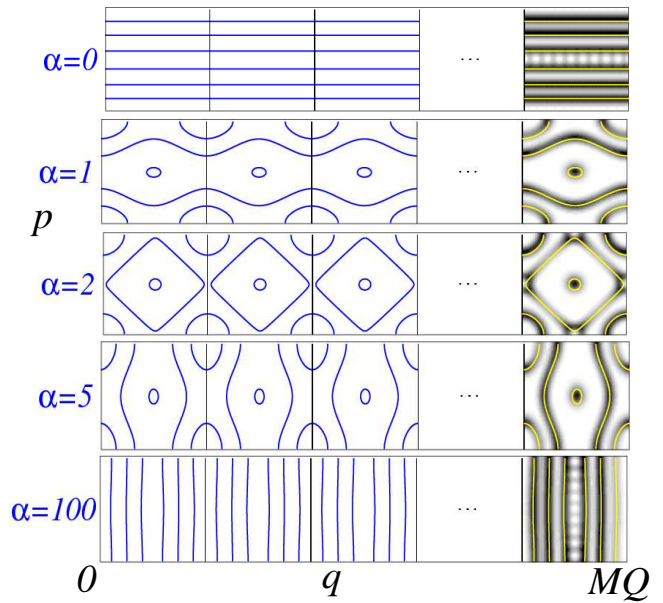}
  \caption{Phase space trajectories of Hamiltonian (\ref{hc}) for different values of the parameter $\alpha=W/J$ (full lines). In the rightest column, the Husimi representation of several eigenstates for $\beta=1/51$ is superimposed to
the classical trajectories (see text).}
\label{psdelta}
\end{figure}

$\mathcal H (q,p)$ defines a time-independent one-dimensional problem which conserves the energy. It is thus an integrable system, with Eq.~(\ref{hc}) defining, for different values of the energy $E={\mathcal H} (q,p)$, one dimensional curves in the phase space plane $(q,p)$. The classical phase-space dynamics is sketched in Fig.~\ref{psdelta} for different values of $\alpha=W/J$. For $\alpha=0$ all trajectories are extended in space (e.g., the projection of the classical curves onto the $q$ axis covers the entire interval $[0,Q]$, in contrast to localized orbits that cover only a fraction of it), and have constant momentum, whereas for $\alpha \gg 1$ exactly the opposite happens, trajectories are localized in space and delocalized in momentum. In between, a crossover is observed, with the presence of two separatrices that delimit three different types of trajectories. The first type are closed localized orbits that oscillate around $(q,p)=(Q/2,P/2)$. This is a standard clockwise oscillation around the minimum of $\cos (2\pi q/Q)$. The second one are also closed localized oscillations, now around $(q,p)=(0,0)$, mod$(Q,P)$, which is a less standard counterclockwise oscillation around the maximum of $\cos (2\pi q/Q)$ due to a local negative mass. Finally, the third type, in between the separatrices, are extended orbits, in $q$ for $\alpha < 2$ and in $p$ for
$\alpha >2$. At $\alpha=2$ the two separatrices merge together. This is the critical value of the parameter $\alpha$ above which all classical trajectories are localized in space.

The quantum mechanical motion is richer than the classical one, because of the additional parameter 
$\beta=\lambda_1/\lambda_2 = 2 \pi\hbar /Q P$. Its presence in the quantum Hamiltonian is a consequence of the discrete nature of the motion in the $q$ direction, defined by the position of the main lattice sites, and the commensurability effects with respect to the secondary lattice. The quantum operator Eq.~(\ref{hh}) commutes with the translation operators $\hat T_Q=\exp(-i Q \hat p/\hbar)$ and $\hat T_P=\exp(i P \hat q/\hbar)$, which translate by one unit cell in each phase-space direction. However, a quantum dynamics simultaneously periodic in both directions, like the classical one, is possible if and only if the translation operators also commute with each other. The most general condition enforces commutation by translations over a phase-space domain of dimensions $(M Q, P)$, where $M$ is an arbitrary positive integer. It is easy to show \cite{leb90} that the operators $T_{M Q}$ and $T_P$ commute if and only if $\beta$ is a rational number
\begin{equation}
\frac{1}{\beta} =  \frac{QP}{2\pi \hbar} = \frac{N}{M} \, ,
  \label{QPnh}
\end{equation}
where $N$ is an arbitrary positive integer. Having in mind the old Weyl rule ``one quantum state per phase-space 
volume $h$'', 
the physical interpretation of this condition is clear: in order to be periodic in both directions, the extended 
phase-space cell $(M Q,P)$, made of $M$ units in the $q$ direction, should contain an integer number $N$ of quantum states. This defines the torus $(M Q,P)$ as the quantum phase space.
An arbitrary quantum state now satisfies the generalized boundary conditions
\begin{equation}
\begin{array}{lll}
T_{M Q} | \psi \rangle &=& e^{i \theta_1} | \psi \rangle \\
T_P | \psi \rangle &=& e^{i \theta_2} | \psi \rangle \ ,
\end{array}
\label{bc}
\end{equation}
where ${\vec \theta}=(\theta_1,\theta_2)$ are good quantum numbers preserved by the dynamics (the Bloch phases). The Hilbert space hence breaks down into $N$ dimensional subspaces parametrized by ${\vec \theta}$. $\hat H$ is therefore a finite-dimensional periodic operator whose spectrum is absolutely continuous. The latter consists of $N$ Bloch bands $E_i ({\vec \theta})$, defined by
\begin{equation}
\hat H | \psi_i ({\vec \theta}) \rangle = E_i ({\vec \theta}) | \psi_i ({\vec \theta}) \rangle
\ , \;\;\;\; i=0,\ldots,N-1 \ ,
  \label{sch}
\end{equation}
where $| \psi_i ({\vec \theta}) \rangle$ are the eigenvectors that satisfy Eqs.~(\ref{bc}). Strictly speaking, 
due to the periodicity, all eigenstates are thus delocalized for any value of the parameter $\alpha$.

Note that the semiclassical limit is obtained when $N \rightarrow \infty$, i.e. when the number of quantum states is large (with, possibly, $M\rightarrow \infty$ but $M/N\rightarrow 0$). In this case the fraction of phase space occupied by one quantum state tends to zero, and the quantum dynamics is described with increasing precision by the classical Hamiltonian, Eq.~(\ref{hc}). The semiclassical limit is, thus, equivalent to the limit $\beta \rightarrow 0$ (or, more generally, and due to the periodicity, to the limit $\beta$ tending to an arbitrary integer from above). In practice, $\beta$ plays the role of an effective Planck's constant. Since $\beta=\lambda_1/\lambda_2$, an appropriate choice of the period of the two superimposed periodic potentials allows to tune $\beta$ and control how quantum or classical the system is. This limit is thus different from the usual semiclassical limit, where one formally takes $\hbar \rightarrow 0$, or considers a high energy regime. From this point of view, there is thus no contradiction in the fact that the classical Hamiltonian $\mathcal H (q,p)$ contains $h$, since the path towards the classical behavior
described by that Hamiltonian is controlled by the two frequencies $\lambda_1$ and $\lambda_2$. This emphasizes that the classical limit $\beta\rightarrow 0$ we are considering has a more general character. It is one example among a wider class of semiclassical limits, sometimes generically called large-$N$ limits, that are often encountered in physics (see for instance \cite{yaffe} and references therein). In our case, the classical limit does not correspond to the standard limit of a classical particle of kinetic energy $p^2 /2m$ moving in the presence of two superimposed periodic potential (cf Eq.~(\ref{h1d}) below). Although purely classical, the structure of $\mathcal H (q,p)$ keeps
information about the original quantum mechanical model, Eq.~(\ref{htb}). The periodicity in the $p$ direction arises from the discretness of the lattice sites distant by $\lambda_1$, imposed by the tight-binding form of the dynamics related to a tunneling process. The boundedness of the kinetic energy term has important consequences, in particular in the appearance of the unusual localized in space-delocalized in momentum classical trajectories observed in the limit
of a large $\alpha$.

Now we turn to the case of irrational $\beta$, whose quantum mechanical behavior deserves special attention. In contrast to our previous discussion, in this case the phase space is not compact, the Bloch angles are not good quantum numbers, and the matrix $\hat H$ is of infinite dimension.  For any irrational value of $\beta$ it can be shown \cite{avila} that the spectrum is a Cantor set (among which the famous Hofstadter butterfly for $\alpha=2$). In addition, the model displays a delocalization-localization transition \cite{aubry,jitomir}. For $\alpha<2$, the spectrum is absolutely continuous and all states are extended. For $\alpha>2$, the spectrum is pure point and all states are localized. Finally, at $\alpha=2$ the spectrum is singular continuous, with multifractal eigenstates \cite{aubry}.

The ideal situation concerning any physical experiment whose aim is to display the delocalization-localization transition is to implement an irrational $\beta$, typically the golden mean $(\sqrt{5}-1)/2$. In practice, one can only approach an irrational as some rational approximation. It is thus of interest to analyze an irrational $\beta$ as the limit of a sequence of rational numbers. The most efficient sequence (in terms of convergence) is known to be the continuous fraction expansion
\begin{equation}
\beta =  \frac{1}{m_1 + \frac{1}{m_2 + \frac{1}{m_3 + \ldots}}} \ .
  \label{cf}
\end{equation}
This sequence of rational approximations $\beta_1 = 1/m_1$, $\beta_2 = m_2/(m_1 m_2 + 1), \ldots$ generates a 
sequence of periodic systems that approximate the quasiperiodic one with increasing accuracy. In the lowest order approximation $\beta_1 = 1/m_1$, only one unit cell is quantized ($M=1$), with $N=m_1$ quantum states supported by this cell. For such a situation, semiclassically the WKB method allows, in 1D, to construct the eigenstates $\psi_i (\vt)$ and eigenvalues $E_i (\vt)$ from the quantization of some specific classical trajectories, in a one-to-one correspondence. The $m_1$ states are selected by the Bohr-Sommerfeld quantization rule \cite{faure}
\begin{equation}
  S=\int_{t} p dq=\beta Q P \left( i+\left[\frac{1}{2}\right]+w_1 \frac{\theta_1}{2\pi}+w_2 \frac{\theta_2}{2\pi}\right) \, ,
  \label{bsqr}
\end{equation}
where $i=0,\ldots,m_1$, $(w_1,w_2)$ are the winding numbers of the quantized classical trajectory $t$ in the $(q,p)$ directions, respectively, and $S=S(E)$ is the action of $t$. The Maslov index $1/2$ applies only to trajectories with $(w_1,w_2)=(0,0)$ (closed trajectories). The eigenvalues of states associated to $(0,0)$ trajectories are thus, at this semiclassical level of approximation, independent of $\vt$.

Eq.~(\ref{bsqr}) associates one quantum state to one classical trajectory of the elementary phase space cell, and defines the spectrum of energies $E_i (\vt)$. In this description, the associated quantum state $\psi_i (\vt)$ can be shown to strongly concentrate or localize around the corresponding quantized classical trajectory.
To illustrate this point, it is useful to display the eigenstates in the phase-space Husimi representation \cite{leb90},
defined in Appendix B. In this representation, to each eigenstate $\psi_i ({\vec \theta})$ of energy $E_i (\vt)$ corresponds a positive-definite phase-space function $W_i (q,p; \vec \theta)$ which, in the semiclassical limit
$1/\beta_1 = m_1 \rightarrow \infty$, behaves as \cite{leb89}
\begin{equation}
W_i (q,p;\vec \theta) \propto \frac{1}{v(q,p)} \exp \left[ - \frac{2}{m_1} \left( 
\frac{{\cal H}(q,p)-E_i (\vec \theta)}{v(q,p)} \right)^2 \right] \ ,
  \label{hus}
\end{equation}
where $v(q,p)=\sqrt{{\dot q}^2 + {\dot p}^2}$ is the phase-space velocity of the classical trajectory of energy $E_i$. The rightest column of Fig. \ref{psdelta} superimposes to the classical trajectories the corresponding quantized eigenstates, for $N=m_1=51$, $M=1$, $\vec \theta = (0,0)$ and different values of $\alpha$. One can clearly see, for a given $\alpha$, the concentration over the corresponding classical trajectory as well as, for varying $\alpha$, the transition from extended to localized states, which simply follow the classical crossover. For open trajectories (either in the $q$ direction for $\alpha<2$ or in $p$ for $\alpha>2$) the quantum state is supported by two symmetric and isoenergetic classical curves, coupled by tunneling. This effect is observed at $\vec \theta = (0,0)$,
where the eigenstates are real. The quantum state combines both classical trajectories, each of which has a current, but whose overall superimposed net current vanishes. As $\vt$ is varied, the eigenstate become complex and can concentrate on one or the other of the classical trajectories.

At the second step of the rational approximation to an irrational $\beta$, $\beta \approx \beta_2 = m_2/(m_1 m_2 + 1)$, the quantum phase-space is an extended torus of dimensions $(m_2 Q,P)$. This space now supports $N=m_1 m_2 + 1$ quantum states. With respect to the previous approximation $\beta\approx\beta_1=1/m_1$, the size of the quantum cell and the number of states $N\approx m_1 m_2$ increase by a factor $m_2$; in contrast the number of quantum states per unit phase-space cell remains almost constant. The scheme repeats as one increases further the order of the continuous fraction approximation. Along this process, the size of the extended phase-space cell increases in the $q$ direction (thus tending towards a cylindrical geometry), the dimension of the corresponding Hilbert space also increases, while the effective Planck constant $\beta$, which measures the fraction of the elementary phase space cell occupied by one quantum state, remains approximately constant. At a given step of the approximation and a given $\vec \theta$, from Eq.~(\ref{bsqr}) it can be shown that two types of quantum states exist, classified according to the structure of the wavefunction in the extended phase space $(M Q, P)$. First, the delocalized ones, which are associated to classical orbits open in the $q$ direction of the $(\pm 1,0)$ type (that exist only for $\alpha < 2$). Second, localized states, associated to either closed orbits (of the $(0,0)$ type) or to open orbits in the $p$ direction (of the $(0,\pm 1)$ type). These states are typically localized over one (or several, when resonances occur) isoenergetic classical orbits of the extended phase space. Their width, controlled by $\beta$, remains almost constant along the continuous fraction approximation. Thus, as the size of the extended torus increases in the $q$ direction, the relative width of the localized states, compared to $M Q$, diminishes. Husimi plots of wavefunctions, that confirm our general picture, 
can be found in Ref.~\cite{aul04} for sequences of rational approximations of the golden mean. 

\begin{figure}
  \includegraphics[width=\linewidth]{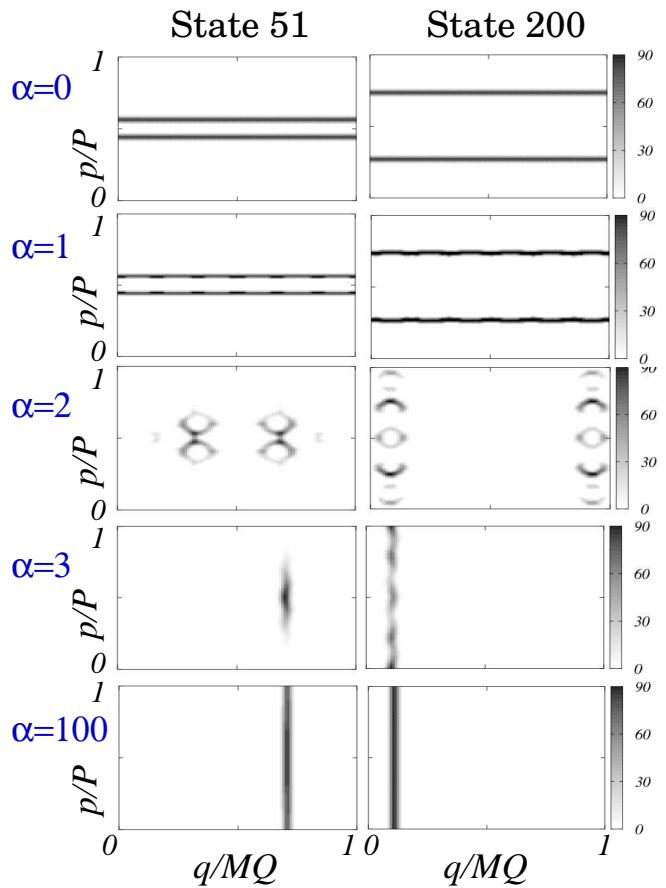}
  \caption{Husimi functions for N=431 and M=516 for two arbitrary quantum states as a function of $\alpha$. Their evolution clearly illustrates the delocalized-localized crossover.}
  \label{husimiexp}
\end{figure}

\section{Relation to experiments}

In recent experiments \cite{inguscio}, a cloud of non-interacting ultracold $^{39}$K atoms created from a Bose-Einstein condensate is released in a bichromatic potential produced by superimposing two lasers of wavelength $\lambda_1$ and $\lambda_2$ (and of corresponding wavenumbers $k_1$ and $k_2$) in a standing wave configuration. The dynamics is described by the following Hamiltonian
\begin{equation}
  \hat H_B = -\frac{\hbar^2}{2m}\frac{\partial^2}{\partial x^2} + V_1 \cos^2(k_1 x) + V_2 \cos^2(k_2 x)\, ,
  \label{h1d}
\end{equation}
where $V_1$ and $V_2$ denote the amplitude of each of the two laser beams and $m$ is the mass of the atom. If the primary lattice (say, $V_1$) is deep enough, that is if the recoil energy $E_R=\hbar^2 k_1^2/2m$ is much lower than $V_1$, one can approximate the motion in the primary lattice by a tunneling process between
localized states of neighboring sites, each of which feels the presence of the superimposed additional periodic
potential $V_2$. The Hamiltonian can thus be reduced to the tight-binding form Eq.~(\ref{htb}), where $\beta=\lambda_1
/\lambda_2$. The effective parameters $J$ and $W$ in Eq.~(\ref{htb}) are directly related to $V_1$, $V_2$ and $E_R$ (see Appendix A).

Several different laser frequencies have been used in the experiments. In \cite{inguscio} they have realized the system described by the Hamiltonian (\ref{h1d}) with $\lambda_1=1032\, nm$ and $\lambda_2=862\,nm$, which yields $\lambda_1/\lambda_2 = 516/431$ (431 is a prime number). Restricting to $0<\beta\leq 1$, we have $\beta = 85/431 \simeq 0.1972$. Moreover, the typical amplitude of the primary lattice is ten times larger than the corresponding recoil energy, which makes (\ref{htb}) a good approximation to the experiment. With the above values of $\lambda_1$, $\lambda_2$ and $\beta$, the quantum system under study is therefore periodic in both $q$ and $p$ directions (because $\beta$ is rational), with $N=431$ states accomodated in $M=85$ elementary cells. This means that every 431 wells of the primary lattice, the system repeats. The Hamiltonian (\ref{htb}) is thus a finite matrix of dimension $N=431$, and the eigenvalues and eigenstates depend parametrically on the two phases ${\vec \theta}=(\theta_1,\theta_2)$. It can be
shown (see Ref.~\cite{leb90} and Appendix B) that the two phases are determine by the relative position of the two lattice sites. Though not directly measured in the previous experiments, the Bloch phases are fixed by the experimental conditions. To be specific, we fix them here to ${\vec \theta}=(0,0)$. Coming back to $\beta$, its continuous fraction decomposition gives two lower order approximants, $\beta_1 = 1/5$, and $\beta_2 = 14/71$.
The system is thus nearly periodic after five wells of the primary lattice. The approximation $\beta \sim \beta_1$ means also that approximately five quantum states are accomodated in one elementary phase space cell $(Q,P)$. Moreover, the relatively low value of $\beta \approx 0.2$ indicates that the classical approximation to the dynamics, given by Eq.~(\ref{hc}), as well as the the semiclassical picture developped in the previous section, is a meaningful framework, as discussed below.

Fig.~\ref{husimiexp} shows the Husimi function of several eigenstates of Eq.~(\ref{htb}) represented in the quantum phase space $(M Q,P)$ for the experimental conditions, namely $\beta=85/431$, and different values of $\alpha$, assuming ${\vec \theta}=(0,0)$. One verifies that the eigenstates follow, for different $\alpha$'s, the general description given at the end of last section. For $\alpha \ll 2$ the eigenstates are concentrated along the delocalized classical trajectories (mixing right and left propagating orbits, because ${\vec \theta}=(0,0)$), while for $\alpha \gg 2$ the eigenstates are concentrated on localized trajectories that wind around in the $p$ direction. In between a crossover between these two limits is observed. This figure illustrates, under the experimental conditions, that the transition from delocalized to localized states as $\alpha$ increases may be interpreted in terms of classical
dynamics, and is therefore different in nature from the Anderson localization. Moreover, as already pointed out in the
introduction, this behaviour is generic and is not restricted to the special set of parameters used in the experiment.
\begin{figure}
  \includegraphics[width=\linewidth]{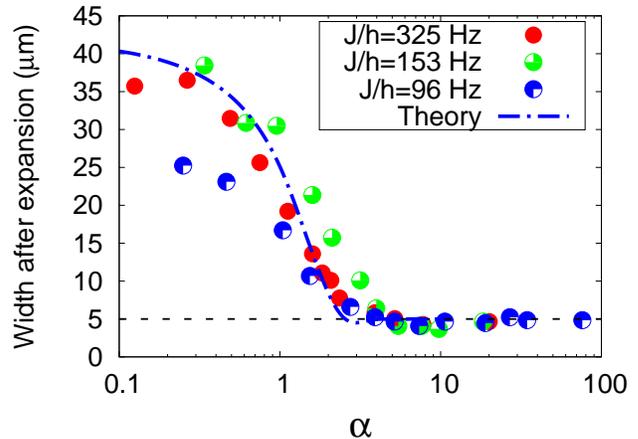}
  \caption{RMS width of a cloud of $^{39}$K atoms after t=750 ms of expansion in a (quasi)periodic potential
  characterized by $\beta = 85/431 = 0.1972 \ldots$ as a function of $\alpha$. The initial width of the cloud is 5  
  $\mu$m, indicated by a horizontal dashed line. Dots : experimental results for different values of $J/h$ (taken from
  Ref.~\cite{inguscio}). Dashed-dot : theory (see text).}
  \label{widthexp}
\end{figure}

Experimentally, the transition was probed through the evolution, for different values of $\alpha$, of an initial cloud of atoms released from the ground state of a confining magnetic trap. The initial state is thus a Gaussian wave packet centered around some initial point $q_0$ and with zero average initial momentum, with an extension of about ten main lattice sites. After letting the cloud expand for some time (750 ms), the final RMS width of the wave packet is measured. Their results are reported in figure \ref{widthexp}. To test our setting, we have computed, using the eigenstates discussed above, the time evolution $|\phi (t) \rangle$ of such an initial state, given by
\begin{equation}
| \phi (t) \rangle = 
\sum_{i=0}^{N-1} \exp \left( -i E_i (\vec \theta) t/\hbar \right) {\overline \psi}_i (z_0)
| \psi_i (\vt) \rangle \ ,
\label{te}
\end{equation}
where, for the actual experiment, $N=431$, ${\overline \psi}_i (z_0)$ is the coherent state representation of the eigenstate $i$, where $z_0 = q_0/\sqrt{2}$ (cf Appendix B), and the Bloch angles ${\vec \theta}$ are set to zero. In Fig.~\ref{widthexp} the width averaged over initial position computed from Eq.(\ref{te}) is compared to the experimental data. The agreement is quite good. Note that the transition takes place here for $\alpha \approx 2$, as predicted by theory, and not around $\alpha \approx 7$, as obtained in Ref.~\cite{inguscio}. The difference is due to an improved estimate of the dependence of the effective parameters $W$ and $J$ upon $V_1$, $V_2$ and $E_R$ (see appendix A and \cite{modugno}).
Note also that the maximal experimental width observed after expansion, of the order of 40$\mu$m, corresponds to
$\sim$80 main lattice sites, much smaller than the phase space cell, that covers 431 main lattice sites.

\section{Comparison with the Anderson model}

In the previous sections we have stressed the classical origin of the delocalization-localization transition, and
of the localized phase, in bichromatic potentials. Let us be more explicit, in particular in the comparison with respect to the Anderson model of localization, by emphasizing their differences and similarities (differences between the quantum motion in disordered and quasiperiodic potentials in terms of effective dispersion relations have been recently analyzed in Ref.~\cite{roux}; see also Ref.~\cite{aul04}).

The original Anderson model of localization \cite{anderson} with diagonal disorder is described, in 1D, by a Hamitonian
similar to Eq.~(\ref{htb}). The difference is that the energy of site $n$, instead of being $\cos (2 \pi \beta n)$, is now given by a random variable $V_n$ having, for instance, a uniform distribution between -1 and 1. Then $W$ simply controls the amplitude of the random potential, and $W V_n$ has a uniform distribution between $-W$ and $W$.

\begin{figure}
\includegraphics[width=\linewidth]{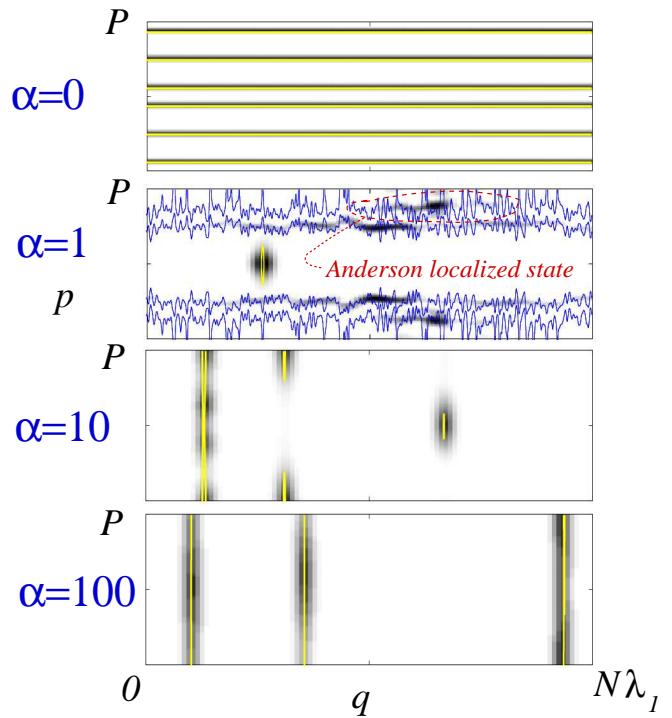}
\caption{Husimi distributions (gray code) of several eigenstates and their corresponding classical orbits (solid lines) in phase space for the Anderson model with $N=401$ and $\sigma=\lambda_1$. As $\alpha$ is increased the classical model also exhibits a delocalized-localized crossover. Localized quantum eigenstates already exist for $\alpha <2$, even if the corresponding classical orbits are delocalized in space, as shown for $\alpha=1$. } 
\label{anderson}
\end{figure}

If one whishes to follow a semiclassical analysis similar to the previous one but for the Anderson model, the first difficulty is to define appropriately the classical limit. The potential, as defined above, is not continuous, and has 
no direct classical limit. One way to get round this difficulty is to smooth the previous random potential, by for instance convoluting $V_n$ with a Gaussian of width $\sigma$
\begin{equation}
V(q) = W \sum_n V_n \frac{e^{-(q-q_n)/2 \sigma^2}}{\sqrt{2\pi \sigma^2}} \ ,
\label{vc}
\end{equation}
where $q_n$ denotes, as in previous sections, the position of the sites of the discrete lattice separated by a distance $\lambda_1$. Then, the classical limit of the Anderson model is described by the Hamiltonian
\begin{equation}
  \mathcal H (q,p) = 2J\cos(2\pi p/P)+ W V(q) \, ,
  \label{hca}
\end{equation}
where $P=h/\lambda_1$. The classical phase space trajectories of such a Hamiltonian are, qualitatively, not very different from those of the quasiperiodic potential Eq.~(\ref{hc}), in particular in the two extreme limits
$\alpha \rightarrow 0$ and $\alpha \rightarrow \infty$ (see Fig.~\ref{anderson}): they are given, again, by delocalized constant-$p$ trajectories in the former limit, and localized, constant-$q$, trajectories in the latter one (we are using, as before, the parameter $\alpha=W/J$). The main difference is in the intermediate regime, where trajectories are more irregularly shaped and a larger number of separatrices is observed. Thus, as in the quasiperiodic Hamiltonian, in the Anderson model a delocalized to localized crossover is observed classically as $\alpha$ increases. 

As we have seen in previous sections, the quasiperiodic case qualitatively follows the classical behavior: a quantum crossover from a delocalized regime to a localized one is observed as $\alpha$ increases. This contrasts with the Anderson model, where states are localized for arbitrary values of $\alpha>0$. Fig.~\ref{anderson} illustrates this point, by showing several quantum states for different values of $\alpha$. We have numerically diagonalized the Anderson model on a finite chain of $N=401$ sites with $\vec \theta=(0,0)$ and obtained the classical phase space orbits from (\ref{hca}) width $\sigma=\lambda_1$. For the sake of clarity Fig.~\ref{anderson} does not show all the classical orbits at a given energy but only the ones directly associated to a quantum state. In the two limiting cases, namely $\alpha=0$ and $\alpha\gg 1$ (here $100$), the quantum states are associated to very simple classical orbits,
i.e. fully delocalized and fully localized trajectories, respectively. However, at finite $\alpha$ the eigenstates are localized, even those whose energy correspond to delocalized classical orbits, as shown in Fig.~\ref{anderson} 
for $\alpha=1$ (of course, the length of the system should be longer than the corresponding localization length).

The main difference between the quantum behavior of the two models is thus in the range $\alpha < 2$, where the Anderson model, in contrast to the classical behavior, has localized states, while the quasiperiodic model follows the classical trajectories. For large values of $\alpha$, the localization in the Anderson model has, as the quasiperiodic potential, a classical interpretation, because the classical orbits are themselves localized. In this respect, the nontrivial regime in the Anderson model is the non-classical localization observed for small values of $\alpha$, an effect which is absent in the bichromatic potential. This stresses the differences and similarities between the two models, in their quantum and classical behavior.

\section{Conclusion}

In the present paper we have analyzed, using semiclassical methods, the delocalized-localized transition
observed in quasiperiodic potentials. A general analysis of the corresponding Aubry-Andr\'e Hamiltonian in 
terms of the commensurability properties of the parameter $\beta$ was given. Particular attention was devoted 
to the interpretation of recent cold atom experiments, as well as to a comparison with the one-dimensional 
Anderson model of localization.

Several extensions and generalizations of the dynamics considered here are of interest, with possible experimental
realizations. They are motivated by the possibility to explore more complex transport effects. Among them, we can
mention the possibility to pulsate in time one of the two superimposed potentials, thus introducing chaotic motion
in the dynamics described by the kicked Harper Hamiltonian \cite{kh}. Along similar lines of research, it is worthwhile to mention recent cold atom experiments concerning the observation of the Anderson metal-insulator
transition \cite{chabe}. Another important effect is related to the role and modifications induced by interactions 
in the delocalization-localization properties in quasiperiodic, as well as random, potentials \cite{int,roux}.

This work was supported by grants ANR--05--Nano--008--02 and ANR--NT05--2--42103 and by the IFRAF Institute.

\appendix

\section{Correction to the Hamiltonian parameters}

In Ref.~\cite{inguscio} they use the following notations. The original Hamiltonian describing the motion of cold atoms in a bichromatic potential is written 
\begin{equation}
  \hat H_B = -\frac{\hbar^2}{2 m}\frac{\partial^2}{\partial x^2} + 
  s_1 E_{R_1} \cos^2(k_1 x) + s_2 E_{R_2}\cos^2(k_2 x)\, ,
  \label{h1ding}
\end{equation}
where $k_i=2\pi/\lambda_i$ and $E_{R_i}=\hbar^2 k_i^2/2 m$, $i=1,2$, denote the wave numbers and
the two recoil energies, respectively. Its reduction to a tight-binding form is expressed as
\begin{equation}
  \begin{array}{lll}
   \hat H &=& \displaystyle J\sum_m | w_{m+1}\rangle \langle w_m| + |w_m\rangle \langle w_{m+1}|\\ \\
     &  + & \displaystyle W \sum_m \cos(2\pi\beta m)|w_m\rangle \langle w_m| \, ,
 \end{array}
  \label{htbing}
\end{equation}
where $\beta=\lambda_1/\lambda_2$ (not restricted to [0,1]). The parameters in (\ref{htbing}), namely $J$ and $W$, were calculated as follows
\begin{equation}
  \begin{array}{l}
   \displaystyle J=1.43 s_1^{0.98}\exp(-2.07\sqrt{s_1})\, ,\\ \\
   \displaystyle W=\frac{s_2 E_{R_2}}{2E_{R_1}}\, ,
  \end{array}
  \label{Ding}
\end{equation}
and thus
\begin{equation}
  \label{a_ing}
  \alpha=\frac{W}{J}=\frac{\beta^2}{2.86} \frac{s_2}{s_1^{0.98}}\exp\left(2.07\sqrt{s_1}\right)\,.
\end{equation}

On the other hand, in Ref.~\cite{boers} the mapping between the two Hamiltonians (\ref{h1ding}) and (\ref{htbing})
have been studied very carefully, in order to analyze the presence of possible mobility edges due to deviations from the tight binding approximation. Writing their formulae with the same set of notations than in Ref.~\cite{inguscio} yields
\begin{equation}
  \alpha=\frac{W}{J}=\frac{\sqrt{\pi}\beta\,^2}{8}
  \frac{s_2}{s_1^{3/4}}\exp\left(2\sqrt{s_1}-\beta\,^2/\sqrt{s_1}\right)\, ,
  \label{a_boers}
\end{equation}
which turns out to present significant deviations with respect to Eq.~(\ref{a_ing}). Indeed, if one takes for instance $s_1=10$ and $\beta=1.2$, which are of the order of magnitude of the experimental parameters in Ref.~\cite{inguscio}, there is about a factor two between the two expressions.

\section{Bargman representation on the torus}

Following \cite{leb90}, we present here the Bargman's representation on the  extended torus. Classically, the two-dimensional toroidal phase space is a periodically repeated cell having sides $(Q,P)$ in suitable coordinates $(q,p)$. The classical dynamics is invariant under translations by the elementary cell. Quantum mechanically, the states of the Hilbert space are required to be periodic functions (up to a phase) under translations defined on an extended torus, of size $(MQ,P)$ ($M$ is an arbitrary strictly positive integer),
\begin{equation}
\begin{array}{lll}
T_{M Q} | \psi \rangle &=& e^{i \theta_1} | \psi \rangle \\
T_P | \psi \rangle &=& e^{i \theta_2} | \psi \rangle \ ,
\end{array}
\label{bca}
\end{equation}
where $\vec \theta = (\theta_1,\theta_2)$ are two Bloch phases ranging from 0 to $2 \pi$, and 
\begin{equation}
\begin{array}{lll}
\hat T_Q=\exp(-i Q \hat p/\hbar) \\
\hat T_P=\exp(i P \hat q/\hbar) \ .
\end{array}
\label{toa}
\end{equation}
To simultaneously satisfy both equations (\ref{bca}), $T_Q$ and $T_P$ must commmute. This imposes that the
area of the extended phase space, measured in units of Planck's constant $h$, must be an integer
\begin{equation}
\frac{M Q P}{2\pi \hbar} = N \, .
\label{weyl}
\end{equation}
The Hilbert space is thus an $N$-dimensional space parametrized by $\vec \theta$, denoted ${\cal H}_N (\vec \theta)$. For a fixed area $MQP$, the semiclassical limit $\hbar \rightarrow 0$ is equivalent to $N \rightarrow \infty$.

For each ${\cal H}_N (\vec \theta)$ one can define normalizable basis states $|q_n, \vt \rangle$ and 
$|p_m,\vt \rangle$ in the $q$ and $p$ representations, respectively
\begin{equation}
\begin{array}{lll}
|q_n, \vt \rangle &=& \sum_{\nu=-\infty}^\infty e^{- i \nu \theta_1} | q_n + \nu M Q \rangle \;\; , n=0,\ldots,N-1 \\
|p_m, \vt \rangle &=& \sum_{\nu=-\infty}^\infty e^{- i \nu \theta_2} | p_m + \nu P \rangle \;\;\;\; , m=0,\ldots,N-1
\end{array}
\label{ba}
\end{equation}
where
\begin{equation}
\begin{array}{lll}
q_n &=& \frac{M Q}{N} (n + \frac{\theta_2}{2 \pi})  \\
p_m &=& \frac{P}{N} (m - \frac{\theta_1}{2 \pi}) \ .
\end{array}
\label{qp}
\end{equation}
In Eq.~(\ref{ba}), $| q_n + \nu M Q \rangle$ and $| p_m + \nu P \rangle$ are the usual position and momentum eigenstates, respectively, that have to be distinguished with respect to their periodized counterpart, 
$|q_n, \vt \rangle$ and $|p_m,\vt \rangle$. The latter states satisfy the boundary conditions Eqs.~(\ref{bca}), and $\langle n,\vt|m,\vt\rangle=\exp(i q_n p_m /\hbar)/\sqrt{N}$. As Eqs.~(\ref{qp}) show, the arbitrariness in the boundary conditions may be viewed as an arbitrariness under shifs of order $1/N$ in the position of the discrete basis states $|n,\vt \rangle$ and $|m,\vt \rangle$ with respect to the intervals $q:[0,MQ[$ and $p:[0,P[$, respectively. 

Diagonalization of the Harper Hamiltonian in, for instance, the $|q_n,\vt \rangle$ basis gives the eigenstates $|\psi_i (\vt) \rangle = \sum_{n=0}^{N-1} \psi_{i,n} (\vt) |q_n, \vt \rangle$, $i=0,\ldots,N-1$, characterized by $N$ complex numbers that satisfy $\sum_{n=0}^{N-1} |\psi_{i,n} (\vt)|^2 =1$. An alternative representation is in terms of coherent states \cite{leb90}, or Bargman representation, defined as 
\begin{equation}
\psi_i (z,\vt) = \sum_{n=0}^{N-1} \psi_{i,n} (\vt) \langle z |q_n,\vt\rangle \ ,
\end{equation}
where
\begin{equation}
\begin{array}{ll}
\langle z|q_n,\vt \rangle =& \left(\pi\hbar\right)^{-1/4} \exp\left(-\frac{1}{\hbar}\left[\frac{1}{2}(z^2+q_n^2)-\sqrt{2}zq_n \right]\right)\\
& \theta_3\left( -\frac{\theta_1}{2}-i\frac{\pi N}{P}(\sqrt{2}z-q_n)\left|\frac{i M N Q}{P}\right.\right) \ .
\end{array}
\end{equation}
The complex variable $z=(q-i p)/\sqrt{2}$ denotes the central position of the coherent state, and
\begin{equation}
\theta_3(v|\tau)=\sum_{\nu=-\infty}^{+\infty}\exp(i\pi\tau \nu^2+2i v \nu)
\end{equation}
is the Jacobi theta function \cite{ww}.

Finally the Husimi representation $W_\psi(z)$ of an eigenstate corresponds to the normalized square
modulus of the coherent state representation
\begin{equation}
  W_i (q,p)=\frac{|\psi_i (z,\vt)|^2}{\langle z|z \rangle}=e^{-|z|^2/\hbar} |\psi_i (z,\vt)|^2\,.
\end{equation}

\end{document}